\documentclass[aps,prd,amsfonts,preprint,
eqsecnum,nofootinbib]
{revtex4}
\usepackage{graphics,epsfig}
\begin{document}
\preprint{WM-03-102}
%
\title{\vspace*{0.3in}
Probing the Light Pseudoscalar Window
\vskip 0.1in}
\author{David L. Anderson}
\email[]{dlande@wm.edu}
\author{Christopher D. Carone}
\email[]{carone@physics.wm.edu}
\author{Marc Sher}
\email[]{sher@physics.wm.edu}
\affiliation{Nuclear and Particle Theory Group, Department of
Physics, College of William and Mary, Williamsburg, VA 23187-8795}
\date{March 2003}
\begin{abstract}
Very light pseudoscalars can arise from the symmetry-breaking sector 
in many extensions of the Standard Model.  If their mass is below $200$ 
MeV, they can be long-lived and have interesting phenomenology.  We 
discuss the experimental constraints on several models with light
pseudoscalars, including one in which the pseudoscalar is naturally 
fermiophobic.  Taking into account the stringent bounds from rare
$K$ and $B$ decays, we find allowed parameter space in each model that 
may be accessible in direct production experiments.  In particular, we
study the photoproduction of light pseudoscalars at Jefferson Lab and
conclude that a beam dump experiment could explore some of the allowed 
parameter space of these models.
\end{abstract}
\maketitle

\section{Introduction}\label{sec:intro}
In many extensions of the standard model, the electroweak
symmetry-breaking sector includes additional weak doublets or
singlets.  New CP-even, CP-odd and charged scalar states may be
present in the physical spectrum. The masses of these particles are
typically of the same order as the weak scale, and fine-tuning is
required to make them much lighter.  An exception occurs if the theory
possesses an approximate global symmetry: a CP-odd scalar may become a
massless goldstone boson in the limit that such a symmetry is exact,
and a massive state that is naturally light in the case where the
symmetry is only approximate.  We will henceforth refer to such CP-odd states as
light pseudoscalars.

The most familiar example of a light pseudoscalar is the
axion~\cite{pq,axion}.  This pseudo-goldstone boson arises in a
two-Higgs-doublet model with a global symmetry that allows independent
phase rotations of the two Higgs fields.  The axion arises as a
consequence of spontaneous symmetry breaking and is exactly massless
in the absence of gauge interactions.  The axion acquires a small mass
due to the QCD anomaly, which breaks this global symmetry at
the quantum level.

In other models, a global symmetry may be broken more significantly by a 
small parameter that appears explicitly in the Lagrangian.  For example, 
consider the Higgs potential for two Higgs doublets~\cite{hhg}, with 
a $\Phi_2\leftrightarrow -\Phi_2$ symmetry:
\begin{equation} 
V= \mu_1^2
 \Phi_1^\dagger\Phi_1 + \mu_2^2 \Phi_2^\dagger\Phi_2 +
 \lambda_1(\Phi_1^\dagger\Phi_1)^2 + \lambda_2
 (\Phi_2^\dagger\Phi_2)^2 + \lambda_3
 \Phi_1^\dagger\Phi_1\Phi_2^\dagger\Phi_2 + \lambda_4
 |\Phi_1^\dagger\Phi_2|^2 +{\lambda_5 \over 2}((\Phi_1^\dagger \Phi_2)^2
+ {\rm h.c.})  
\label{eq:hpot}
\end{equation} 
In the limit $\lambda_5\rightarrow 0$, this potential has a $U(1)
\times U(1)$ symmetry in which each doublet rotates by an independent
phase.  The spontaneous breaking of the diagonal U(1)
symmetry yields a goldstone boson that is ``eaten'' when the theory is
gauged; the remaining U(1), which rotates each doublet by an opposite
phase, yields a physical goldstone boson state.  When $\lambda_5$ is
nonvanishing, this pseudoscalar  develops a mass given by
$m^2_A=-\lambda_5v^2$, where $v=246$ GeV is the electroweak scale.  In
this paper, we will consider pseudoscalars with masses in the
$100-200$ MeV range, for phenomenological reasons explained below.
This can be achieved by setting $\lambda_5$ equal to a small number
that is comparable to a light fermion Yukawa coupling---a light
pseudoscalar would then be no more or less unnatural than a muon or
light quark.

Of course, one can construct models in which the light fermion Yukawa
couplings arise only via higher-dimensions operators in a more complete
high-energy theory.  The Yukawa couplings are identified with powers of
the ratio of a symmetry breaking scale to the cut off of the theory, and
therefore can be naturally small. By analogy, the U(1) symmetry
present in the $\lambda_5=0$ limit of Eq.~(\ref{eq:hpot}) may be
broken by a field $\eta$ that acquires a vacuum expectation value at
some high scale and contributes to the term of interest only through
Planck-suppressed operators.  Given this dynamical assumption, one
predicts that the pseudoscalar mass is of the order $(\langle \eta
\rangle/ M_*)^{n/2}\, v$, where $n$ is a positive integer, and $M_* =
2 \times 10^{18}$~GeV is the reduced Planck mass.  Interestingly, for
$n=2$, and $\langle \eta \rangle\sim 10^{15}$~GeV (the
nonsupersymmetric GUT scale), one obtains a pseudoscalar mass of
approximately $100$~MeV.  One can imagine a variety of high energy
theories in which similar results are obtained.

Our interest in pseudoscalar masses between $100$ and $200$~MeV is
motivated by the pseudoscalar decay length and production cross
section.  We hope to have both in optimal ranges for detection of the
pseudoscalar in possible photoproduction experiments at Jefferson Lab.
As far as production is concerned, existing direct searches yield
bounds on the pseudoscalar couplings that are weakest in this mass
range, and a wide variety of
experiments~\cite{dawson,sundrum,direct,hhg} severely constrain the
pseudoscalar couplings for masses below $100$~MeV.  On the other hand,
if the pseudoscalars are produced in significant but not overwhelming
numbers, we hope for a decay length that is long enough to clearly
separate the pseudoscalar decay signal from possible mesonic
backgrounds. Pseudoscalars with masses above $200$~MeV decay rapidly
into muon pairs with a branching fraction near $100\%$, making
detection via a separated vertex impossible. Thus, the $100-200$~MeV
mass window seems particularly promising for the experimental search
that we propose in Section IV. 

To proceed with our phenomenological analysis, we must decide on the
pseudoscalar's couplings to standard model fermions; the pattern of
these couplings is in fact quite model-dependent. In the standard
two-Higgs-doublet models, the pseudoscalar couplings are proportional
to Yukawa matrices multiplied by a ratio of the vacuum expectation
values $v_1$ and $v_2$.  On the other hand, one can employ simple
discrete symmetries to construct three-doublet models in which only
two doublets couple to quarks and do not mix with a third doublet
coupling to the leptons.  In this case, the pseudoscalar in the
quark-two-doublet sector is entirely leptophobic.  An analogous
three-doublet model with a lepton-two-doublet sector yields a
pseudoscalar that has no couplings to quarks and is, hence,
hadrophobic.  Such models illustrate the range of the possible, but
are not particularly well motivated.  A much more appealing
possibility is that the pseudoscalar may have no direct couplings to
quarks or leptons at all.  Let us comment on the motivation for such a
fermiophobic pseudoscalar in more detail.

One could imagine a number of reasons why a pseudoscalar may have
suppressed couplings to standard model fermions.  The suppression
could be {\em parametric}, as in the type-I two Higgs doublet model
when $\tan\beta$ is taken large. On the other hand, the suppression
could be {\em geometric}, as in extra-dimensional scenarios in which
fields have wave functions that are localized at different points in
an extra dimension.  Let us focus on a concrete realization of this
second idea. Consider an $S^1/Z_2$ orbifold of radius $R$, with
standard model matter fields located at the $y=0$ fixed point, and
gauge fields in the 5D bulk. Here $y$ is the extra-dimensional
coordinate.  Assume that there exists additional vector-like matter in
complete SU(5) representations (to preserve gauge coupling
unification) as well as a gauge-singlet scalar field $S$, all isolated
at the $y=\pi R$ fixed point.  A spontaneously broken approximate
global symmetry of the singlet potential leads to a light pseudoscalar
state that couples directly to the exotic matter multiplets only.  The
geometry of this scenario prevents mixing between the ordinary and
exotic matter fields, which communicate with each other only via gauge
interactions in the bulk.  The scale of compactification can be taken
large enough so that the effects of Kaluza-Klein excitations are
irrelevant to the low-energy theory.  

Given the simplicity of the fermiophobic singlet scenario described
above, we will focus our discussion on light pseudoscalars in the
two-Higgs-doublet models of type-I and II and in the fermiophobic
singlet scenario.  We comment on the other possibilities where
appropriate.  In Section II, we analyze the experimental constraints
on the light pseudoscalar in the conventional two-Higgs-doublet
models, placing particular emphasis on the bounds from $K$ and $B$
meson decays. In Section III, the fermiophobic singlet scenario is
studied, and in Section IV we study the possibility of detecting
pseudoscalars of either type in photoproduction experiments at
Jefferson Lab.  Section V contains our conclusions.

\section{Constraints in Two-Doublet Models}

As we have described in the previous section, light pseudoscalars can
arise in two-Higgs-doublet extensions of the standard model.  Two
popular options exist in which a discrete symmetry is imposed to
forbid tree-level flavor changing neutral
currents~\cite{glashowweinberg}: In Model I, all of the fermions
couple to a single Higgs doublet, but none to a second.  In Model II,
the charge $Q=2/3$ quarks couple to one Higgs doublet while the
$Q=-1/3$ quarks and the leptons couple to another.  A third
possibility is that all fermions couple to both Higgs doublets,
without the restriction of any discrete symmetry.  An ansatz is then
employed to make tree-level flavor changing Higgs couplings
sufficiently small~\cite{chengsher}.  However, in this case it has
been shown that a very light pseudoscalar will still lead to
unacceptably large flavor-changing neutral currents~\cite{sheryuan}.

The coupling of the pseudoscalar Higgs to fermions is of the form
$-{m_f\over v}X_f\bar{f}\gamma_5fA$ where $v=246$ GeV and
$X_f=\cot\beta$ for all fermions in Model I, and $X_f=\cot\beta\
(\tan\beta)$ for the $Q=2/3$ quarks ($Q=-1/3$ quarks and leptons) in
Model II.  Here $\tan\beta$ is the ratio of vacuum expectation values
of the two Higgs doublets, and is a free parameter.

There have been numerous discussions of the bounds on a light
pseudoscalar, most recently by Larios, Tavares-Velasco and
Yuan~\cite{cpyuan}.  In Model II, the combined bounds from the
nonobservation of $J/\Psi\rightarrow A\gamma$ and $\Upsilon\rightarrow
A\gamma$ force $\tan\beta$ to be close to $1$, since the former decay
implies $\tan\beta\lesssim 1$ and the latter implies
$\cot\beta\lesssim 1$~\cite{hhg}; theoretical uncertainties don't
quite allow the model to be excluded.  In Model I, both decays imply
only that $\cot\beta\lesssim 1$.  Bounds from $\eta$, $\eta^\prime$
and $\pi$ decays also force $\tan\beta\sim 1$ in Model II and
$\cot\beta\lesssim 1$ in Model~I~\cite{pich}.  Bounds from $g-2$ are
in flux at the moment, but do not appreciably change these
results. (In addition, the $g-2$ bound is only valid if one makes a
strong assumption that there are no other possible nonstandard
contributions at one loop.)  Bounds from $b\rightarrow s\gamma$,
$\Delta\rho$, $R_b$ and $A_b$ can all be avoided by constraining the
neutral and charged scalar masses~\cite{cpyuan}.  Thus, we will
consider two cases: Model II with $\tan\beta\sim 1$ and Model I with
$\cot\beta\lesssim 1$.  After reviewing the decay modes and decay
lengths of the light pseudoscalar, we consider the bounds from $K$ and
$B$ meson decays, which present the strongest constraints on these
models.

\subsection{Decay Modes}

For a pseudoscalar lighter than twice the muon mass, there are only
two possible decay modes, $A\rightarrow e^+e^-$ and $A\rightarrow
\gamma\gamma$.  The decay width into an electron pair is given by
\begin{equation} 
\Gamma_{A\rightarrow e^+e^-} = {m_e^2\over 8\pi v^2}M_A X_e^2\left(
1-4{m^2_e\over M^2_A}\right)^{1/2} \,\, .
\end{equation} 
For $\tan\beta=1$, this gives a decay length of $0.6-1.2$ centimeters
in the pseudoscalar rest frame, for $M_A$ ranging from $100$ to $200$
MeV. This result scales as $\tan^2\beta$ in Model II and $\cot^2\beta$
in Model I.

The decay into two photons proceeds at one loop with the width
\begin{equation}
\Gamma_{A\rightarrow \gamma\gamma} = {|\sum_f N_cQ^2_fX_f|^2 \alpha^2
M_A^3 \over 64\pi^3v^2} \,\, ,
\label{eq:twophot}
\end{equation} 
where $N_c$ is $3$ for quarks and $1$ for leptons and $Q_f$ is the
fermion charge.  This expression is valid if the mass of the fermion
in the loop is much larger than the momentum in the decay.  When this
is not the case then the exact expression given in
Refs.~\cite{cpyuan,hhg} should be used.  Note that
Eq.~(\ref{eq:twophot}) is independent of the heavy fermion mass.  For
the top quark contribution alone, with $\tan\beta=1$, one obtains a
decay length in the pseudoscalar rest frame of $30$ centimeters for
$m_A=100$ MeV.  Note that if one considers all quarks and leptons
except the first generation fields, then the decay width is increased
by a factor of $16$, which would correspond to a decay length of 2
centimeters.  For $\tan\beta\sim 1$, the branching ratio into photons
is $10\%$ for $M_A=100$ MeV and $40\%$ for $M_A=200$ MeV.  Thus, we
see that typical decay lengths, for $\tan\beta=1$, are on the order of
a centimeter.  For Model I with small $\cot\beta$, this decay length
is increased by a factor of $\tan^2\beta$.  These decay lengths will,
of course, be increased by a relativistic factor if the pseudoscalar
has a large momentum (as it does in $B$-decays).

\subsection{K decays}

It is has been long known that the strongest bounds on axion models
come from the decay $K\rightarrow \pi A$~\cite{hallwise,frere}; one
expects that the same process will significantly constrain the light
pseudoscalar scenarios of interest to us here.  While many early
analyses (that did not take into account the heaviness of the top
quark) seemed to exclude the possibility of a light pseudoscalar in
the standard two-doublet scenarios, more recent work suggests that an
allowed window remains.  It was pointed out by Grzadkowski and
Pawelczyk that there are two contributions to the decay amplitude and that
the sum may vanish for some choices of model parameters~\cite{grz}. The first
is a direct decay contribution involving the top quark and charged
Higgs bosons at one loop; the second is an indirect contribution
following from mixing between the axion and the $\pi^0$, $\eta$ and
the $\eta'$.
\begin{figure}[ht]
\epsfxsize 3.3 in \epsfbox{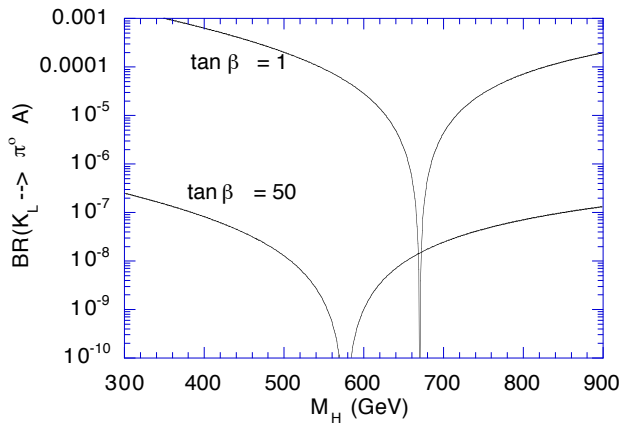}
\caption{The branching ratio for $K_L\rightarrow \pi^0 A$ for two values of
$\tan\beta$ as a function of the charged Higgs mass.  We choose
$M_A=150$ MeV.  The experimental bound is approximately $4\times
10^{-8}$.}
\label{klong}
\end{figure}
We refer the reader to Ref. \cite{grz} for the full expressions. As an example,
the amplitude for $K^+\rightarrow \pi^+ A$ in Model I can be
written schematically as
\begin{equation}
\lambda_w\cot\beta\ F(m_K,m_\pi,m_A,m_\eta,m_{\eta^\prime})
+ \cot\beta\ G(\beta,m_{top},m_{H^+},U_{CKM}) \,\, .
\label{eq:schematic}
\end{equation}
The first term depends only on meson masses and is due to the
pseudoscalar mixing; $\lambda_w$ is a chiral Lagrangian parameter that
is fixed by the data to be $|\lambda_w|=3.2\times
10^{-7}$~\cite{manohar}.  The sign of $\lambda$ can be determined by
matching chiral Lagrangian amplitudes to electroweak
results~\cite{hhg} and is negative (the imaginary part is proportional
to the CP violating factor $\epsilon$ \cite{goity} and is thus
negligible).  The second term represents the direct, one-loop decay
amplitude, and depends on the top mass, the charged Higgs mass, and on
CKM angles.  Specifically, the second term may be written
\begin{equation}
-{1\over 2}(m^2_{\pi}-m_K^2){\xi\over v}
\end{equation}
where
\begin{equation}
\xi=-{G_F\over 16\pi^2}\sum_q U_{qs}U_{qd}^*m_q^2\cot\beta
(A_1+\cot^2\beta A_2)  \,\,\, .
\label{eq:a1a2}
\end{equation}
Here $A_1$ and $A_2$ are functions of the top, charged Higgs and W
masses and are given explicitly in Ref.~\cite{frere}.  Numerically,
the first term of Eq.~(\ref{eq:a1a2}) is typically a few times
$10^{-11}$~GeV and the second is typically $10^{-9}$~GeV.
However, Grzadkowski and Pawelczyk show that the second term changes
sign as the charged Higgs mass varies from $50$ GeV to $1000$ GeV, and
thus at some value the total amplitude vanishes.  We have plotted
their results for the $K_L$ decay in Fig.~1, setting $\tan\beta=1$
(so our results then apply to both Model I and Model II), and also
$\tan\beta=50$ in Model I.  Consideration of $K^+$ and $K_S$ decays
leads to qualitatively similar results.
\begin{table} [ht]
\center{
\begin{tabular}{ccccccccc} \hline\hline
$\tan \beta$ &\hspace{1em} & $K_S$ &
\hspace{1em} & $K_L$ &\hspace{1em} & $K^{\pm}$ & \hspace{1em} & $B$\\ \hline
1 & & 661-693 & & 668-672 & & 669-672 & & 662-678 \\
2 & & 576-643 & & 597-607 & & 599-606 & & 599-605 \\
3 & & 546-648 & & 580-596 & & 583-594 & & 584-592 \\
4 & & 526-662 & & 572-594 & & 576-591 & & 578-588 \\
5 & & 508-679 & & 567-595 & & 571-591 & & 575-587 \\
10 & & 434-781 & & 550-607 & & 558-599 & & 566-590 \\
15 & & 371-900 & & 536-621 & & 548-609 & & 560-595 \\
20 & & 317-1036 & & 522-637 & & 538-620 & & 554-601 \\
30 & & 227-1369 & & 496-669 & & 518-642 & & 542-614 \\
40 & & 158-1804 & & 472-702 & & 500-665 & & 531-626 \\
50 & & 105-2370 & & 448-738 & & 482-689 & & 520-639 \\ \hline\hline
\end{tabular} }
\caption{The allowed ranges for the charged Higgs mass (in
GeV) for $K_S$, $K_L$, $K^{\pm}$, and $B$ decays.  The four ranges 
overlap for all $\tan\beta$ shown.\label{table1}}
\end{table}

>From Fig.~\ref{klong} we see that there is a very narrow region of
parameter space in which the branching ratio is suppressed.  We now
must consider whether the experimental bounds on $K_S$, $K_L$ and
$K^\pm$ decays can be satisfied simultaneously. The Higgs Hunters
Guide~\cite{hhg} refers to two experiments~\cite{kone,ktwo} that
search for the decay chain $K^+\rightarrow\pi^+ A$,~$A\rightarrow
e^+e^-$, and obtain upper limits on the $\pi A$ branching ratio of
order $10^{-8}$.  However, it is important to point out that a region
between $m_A=100-150$~MeV remains unconstrained due to the large
background from the standard decay $K^+\rightarrow \pi^+\pi^0$,
followed by $\pi^0$ Dalitz decays.  Without precise vertex detection,
this can not be distinguished from the pseudoscalar signal.  In the
particular case of Model I with large $\tan\beta$, the decay length
increases by $\tan^2\beta$, and can be several meters.  The pseudoscalar
would then escape the detector.  In that event, bounds from
$K^+\rightarrow \pi^+ \,\, nothing$~\cite{kthree,kfour,bazarko}, which 
range from $10^{-7}$ to $10^{-10}$, would apply.  Again, the weaker
${\cal O}(10^{-7})$ bound applies to a mass interval between 
$m_A=130-160$~MeV, as a consequence of larger experimental backgrounds.
On the other hand, the experimental bounds on the decay $K_L\rightarrow \pi^0 A$, 
are uniformly strong over the entire range of pseudoscalar masses~\cite{klongdecay}.
Fortunately, one can fine-tune the charged Higgs mass to avoid
contradiction with both charged or neutral kaon decay bounds.  In Table I, we 
show the required range of charged Higgs masses for $K^+$, $K_L$ and $K_S$ decays. 
It has been assumed that the $A$ mass is $150$ MeV, so that the tighter experimental
bounds in charged K decays apply; if the mass is between $100$ MeV and
$150$ MeV, these bounds are relaxed and the ranges for $K^+$ and $K_S$
decays are much wider.  For all values of $\tan\beta$ shown in
Table~\ref{table1}, the allowed ranges for charged Higgs mass overlap
and all the bounds can be satisfied with a single fine tuning.  For
Model II, in which $\tan\beta\sim 1$, the charged Higgs mass must be
tuned to approximately one percent precision, but in Model I with
larger $\tan\beta$, relatively mild fine-tuning is sufficient.  Thus,
kaon decays cannot completely exclude the existence of a pseudoscalar
in the $100-200$ MeV mass range.

\subsection{B decays}

In B decays into $K A$, the pseudoscalar will have a relativistic
gamma factor of $12-24$, depending on its rest mass.  Thus, the decay
length into electrons will be approximately $25$ centimeters (times
$\tan^2\beta$).  Because of the larger CKM mixing  with the top quark,
the Higgs-top loop contribution to the amplitude generally
dominates over the mixing term by a larger amount than in
the case of kaons.  A simple estimate illustrates that the
branching fraction is potentially large:  The loop term involves
CKM factors that are comparable to those found in tree-level semileptonic
decays, while the $16\pi^2$ in the loop is partly compensated by the smaller 
two-body phase space.  The resulting prediction has a shape very
similar to that for $K$ decays in Fig.~1.  Again, there is a narrow
region of parameter-space where the rate vanishes, and this region
matches the narrow region in K-decays.  This is not surprising since
the analog of Eq.~(\ref{eq:a1a2}) for $B$ decays has the same functional 
dependence on the charged Higgs mass, up to an overall factor.  One might 
hope that higher order corrections would separate the $K$ and $B$ decay
allowed mass windows, but a one-percent effect would not be
sufficient to alter our qualitative results.

What are the experimental limits?  Recently, the BELLE Collaboration
published a value for the branching fraction for $B\rightarrow
Ke^+e^-$ of $0.75\pm 0.2\ \times 10^{-6}$~\cite{belle}.  Since this is
in agreement with theory, a bound on new physics contributions of
approximately $2\times 10^{-7}$ can be obtained.  However, the BELLE
analysis included a mass cut on the electron-positron pair of $140$
MeV, to suppress background from photon conversions and $\pi^0$ Dalitz
decays.  Thus, the bound does not apply to the $100-140$ MeV window.
The CLEO Collaboration has searched for $B^\pm \rightarrow K^\pm \,\,
nothing$ and $B^0 \rightarrow K^0_S \,\, nothing$ decays, and obtains
a bound on the branching ratios of $5\times 10^{-5}$~\cite{urheim}.
While this does cover the mass range in which the BELLE analysis does
not apply, it is only relevant if all the pseudoscalars escape
detection. For masses between $100$ and $140$ MeV, one can ask what
fraction of the $A$'s will escape the detector.  For $\tan\beta=50$,
the decay length will be over 10 meters and almost all of the $A$'s
would escape; the CLEO bound would then apply.  In general,
approximately $e^{-4\cot^2\beta}$ of the $A$'s escape the detector,
which is a barrel calorimeter of roughly a meter radius.  The bound
would then be weaker by this factor, or $5\times
10^{-5}e^{4\cot^2\beta}$ for the branching ratio.  Using this
experimental bound, we find the allowed charged Higgs mass range given
in Table I.  We see that the same fine-tuning needed (for
$\tan\beta\sim 1$) for kaon decays will automatically suppress the
B-decay rate.

We conclude that neither model I nor II can be definitively excluded from
the bounds from $B$ decay, although fine-tuning is needed if 
$\tan\beta\sim 1$, as required in Model II.

\subsection{Leptophobic Pseudoscalars}

As noted in the introduction, it is simple to have a three Higgs model
in which two of the Higgs doublets couple to quarks (with Model I or
Model II couplings) and a third couples to leptons.  If the third
doublet does not mix with the others, the leptonic couplings of the
light pseudoscalar are eliminated. The $K$ and $B$ decays discussed in
the previous two subsections will generally not be affected in such a
model. However, the decay of the pseudoscalar will now be entirely
into photon pairs and the lifetime will generally be 2-3 times larger
than the usual case.  Note that in the $100-140$~GeV mass window, the 
stronger bounds from $K_L$ decays and from CLEO will certainly apply
(without any significant exponential correction for decays inside the
detector).  Again, these bounds can be evaded with a suitable fine
tuning of the charged Higgs mass.

\section{Fermiophobic Pseudoscalars}
   
We have seen in the previous sections that a pseudoscalar state in the
$100$ to $200$~MeV mass range is consistent with the stringent bounds
from $K$ and $B$ meson decays.  However, in the conventional scenarios
considered thus far, this result follows from an accidental zero in
the decay amplitudes, as well as a willingness to accept fine tuning.
In this section we consider another possibility, that the couplings of
the pseudoscalar to matter are naturally suppressed.  After discussing
the experimental bounds, we argue that a natural place to search for
such a state is in a low-energy photoproduction experiment, such as
those possible at Jefferson Lab.  We estimate the production rate and
comment on the relevant discovery signal in Section IV.

We have already stated the motivation for considering a pseudoscalar
state that is light: it might be the would-be goldstone boson
associated with a global symmetry that is only approximate.  In the
introduction, we outlined a plausible scenario with a singlet scalar
and a vectorlike multiplet in a complete $SU(5)$ representation, taken
to be a $5+\bar{5}$ for simplicity.

Since the exotic matter is vector-like, it can be made arbitrarily
heavy and integrated out of the theory. This leads to
nonrenormalizable interactions between the pseudoscalar and the
standard model gauge fields.  If $M_F$ is the mass scale of the
vector-like matter $\psi$, and the pseudoscalar coupling is given by
$(i A \lambda/\sqrt{2}) \bar\psi \gamma^5 \psi$, then one obtains
\begin{equation}
{\cal L} = \frac{q^2 \lambda}{32 \sqrt{2} \pi^2
M_F}\epsilon^{\mu\nu\rho\sigma}\, A F_{\mu\nu} F_{\rho\sigma}
\label{eq:effop}
\end{equation}
for the effective coupling of the pseudoscalar to two photons. Here
$q$ represents the electric charge of $\psi$, and $F_{\mu\nu}$ is the
electromagnetic field strength.  Note that this can be generalized to
any non-Abelian gauge group by replacing $q^2$ with the Casimir $T_F$
(defined by $\mbox{Tr}[T^a T^b]= T_F \delta^{ab}$) and by summing over
the field strengh tensors.  For a fermiophobic pseudoscalar in the
mass range of interest to us, the only possible decay is to two
photons, and from Eq.~(\ref{eq:effop}) we obtain the decay width
\begin{equation}
\Gamma(A\rightarrow \gamma\gamma) = \frac{16}{9}\cdot \frac{\alpha^2
\lambda^2}{128 \pi^3} \frac{m_A^3}{M_F^2} \,\,\, .
\label{eq:aggwidth}
\end{equation}
If $M_F$ is not far above the top quark mass, say $200$~GeV, and
$\lambda=1$, then one obtains a lifetime
\begin{equation}
\tau (M_f=200\mbox{ GeV}) = 1.1 \times 10^{-3}\mbox{ sec } \left(
\frac{\mbox{MeV}}{m_a}\right)^3 \,\,\, .
\end{equation}
For energies of a few GeV, typical of the photoproduction experiments
that we will mention later, the pseudoscalar can travel a macroscopic
distance before it decays.  A pseudoscalar with a mass of $150$ MeV
and an energy of $3$ GeV will have a decay distance of $160$
centimeters.

One might think that the scenario described above is relatively
insensitive to the bounds from meson decays due to the weakness of the
pseudoscalar's coupling to ordinary matter.  However, the experimental
bounds on the branching fraction of $K$ or $B$ mesons to
$\pi+$~pseudoscalar are so stringent that operators like
Eq.~(\ref{eq:effop}) are potentially significant, even when they
contribute only at one loop.  Here we estimate the contribution to $K
\rightarrow \pi A$ in order to constrain the parameter space of the
model.  We comment on the constraints from $B$ decays at the end of
this section.

The operator with the largest potential effect on low-energy
hadronic decays is the gluonic version of Eq,~(\ref{eq:effop}).  We
use a chiral lagrangian approach to estimate the branching fraction of
interest~\cite{grz}.  First we represent the light pseudoscalar nonet via the
nonlinear representation
\begin{equation}
\Sigma = \exp(2i \pi/f_\pi)
\end{equation}
where $f_\pi=93$~MeV is the pion decay constant, and where $\pi$ is
the matrix of fields
\begin{equation}
\pi = \left(\begin{array}{ccc}
\frac{\pi^0}{2}+\frac{\eta}{2\sqrt{3}}+\frac{\eta'}{\sqrt{6}} &
\frac{\pi^+}{\sqrt{2}} & \frac{K^+}{\sqrt{2}} \\
\frac{\pi^-}{\sqrt{2}} &
-\frac{\pi^0}{2}+\frac{\eta}{2\sqrt{3}}+\frac{\eta'}{\sqrt{6}} &
\frac{1}{2} (K^0_s+K^0_L) \\ \frac{K^-}{\sqrt{2}} & \frac{1}{2}
(K^0_L-K^0_s) & -\frac{\eta}{\sqrt{3}}+\frac{\eta'}{\sqrt{6}}
\end{array}\right) 
\end{equation}
Here we have ignored CP violation and expressed the neutral kaons in
terms of their CP eigenstates.  Also note that we have chosen to
include the $\eta'$, so that $\Sigma$ is an element of U(3) rather
than SU(3).  The $\Sigma$ field transforms simply under the chiral
SU(3) symmetry
\begin{equation}
\Sigma \rightarrow U_L^\dagger \Sigma U_R
\end{equation}
leading to the usual lowest order effective Lagrangian
\begin{equation}
{\cal L}_0 = \frac{f^2}{4} \mbox{Tr } \partial_\mu \Sigma^\dagger
\partial^\mu \Sigma +\frac{1}{2} f^2 \mu \mbox{Tr }(M \Sigma^\dagger +
\Sigma M^\dagger) \,\,\, ,
\label{eq:lzero}
\end{equation}
where $M$ represents the light quark current mass matrix.  However,
Eq.~(\ref{eq:lzero}) does not take into account the QCD anomaly, which
relates the divergence of the axial current to the product of gluon
field strength tensors $G^{\mu\nu}\tilde{G}_{\mu\nu}$.  A possible
method of incorporating this effect into the chiral lagrangian is to
introduce the additional terms~\cite{chiralanom}
\begin{equation}
{\cal L}_{anom} = \frac{1}{2} i q(x)
\log\frac{\det\Sigma}{\det\Sigma^\dagger} + c q(x)^2
\label{eq:anoml}
\end{equation}
where $q(x)$ represents
\begin{equation}
q(x) = \frac{g^2}{32\pi^2} G_{\mu\nu}^a \tilde{G}_a^{\mu\nu} \,\,\, .
\end{equation}
Under an axial U(1) rotation, the field $\Sigma$ is multiplied by an
overall phase, and it is not hard to show that ${\cal L}_0 + {\cal
L}_{anom}$ yields the appropriate divergence of the axial vector
current~\cite{chiralanom}.  Now, one may treat $q(x)$ as an auxilliary 
``glueball'' field, and remove it using its equation of motion.  One then finds
\begin{equation}
{\cal L}_{anom}=-\frac{1}{4c} \left( \frac{i}{2} \log \frac{\det \Sigma}
{\det \Sigma^\dagger}\right)^2 \,\,\, .
\label{eq:almost}
\end{equation}
This term determines the $\eta'$ mass, and the parameter $c$ can be chosen
accordingly.  If one now includes the pseudoscalar coupling to gluons, an
additional term must be added to Eq.~(\ref{eq:anoml}), namely
$A q(x)/(2\sqrt{2}M_F)$, in which case Eq.~(\ref{eq:almost}) is modified
\begin{equation}
{\cal L}_{anom}=-\frac{1}{4c} \left( \frac{i}{2} \log \frac{\det \Sigma}
{\det \Sigma^\dagger} + \frac{1}{2\sqrt{2}M_F} A \right)^2 \,\,\, .
\end{equation}
This interaction leads to mass mixing between the pseudoscalar and the $\eta'$;
we find that the mixing angle is given approximately by
\begin{equation}
\theta_{A\eta'} \approx \frac{1}{4\sqrt{3}}\frac{f_\pi}{M_F}  \,\,\,
\end{equation}
or numerically, $7 \times 10^{-5}\cdot (200 \mbox{ GeV}/M_F)$. We may extract
the $\Delta S=1$ $K\pi\eta'$ vertex from the chiral Lagrangian term
\begin{equation}
{\cal L}_{\Delta S=1} = \frac{f_\pi^2}{4} \mbox{Tr }(\lambda_w h \partial_\mu
\Sigma \partial^\mu \Sigma^\dagger)
\end{equation}
where $h$ is octet-dominant $\Delta S = 1$ spurion
\begin{equation}
h = \left(\begin{array}{ccc} 0 & 0 & 0 \\ 0 & 0 & 1 \\ 0 & 0
& 0
\end{array}\right)  \,\,\, ,
\end{equation}
and $\lambda_w=3.2 \times 10^{-7}$ is a parameter that takes into account the
strength of the weak interactions~\cite{manohar}.  We find
\begin{equation}
\Gamma(K^+ \rightarrow \pi^+ A) = \frac{1}{384\pi}\,\frac{\lambda_w^2
\, \theta_{A\eta'}^2}{m_K^3 f_\pi^2} \,(m_A^2+2 m_K^2)^2
\,[(m_K^2-m_\pi^2+m_A^2)^2-4 m_K^2 m_A^2]^{1/2} \,\, .
\end{equation}
As a point of reference, if one sets $m_A=100$~MeV, one obtains the
branching fraction $5.6 \times 10^{-7} \cdot (200 \mbox{ GeV}/M_F)^2$.

Different experimental bounds are relevant depending on the lifetime
and boost of the pseudoscalar. If the pseudoscalar decays inside the
experimental detector, the relevant bound on the $K^+$ branching
fraction is~\cite{hhg}
\begin{equation}
BF(K^+ \rightarrow \pi^+ \gamma \gamma) < 1.4 \times 10^{-6} \,\,\, .
\label{kbound1}
\end{equation}
If the pseudoscalar escapes the detector unobserved, one must contend
with more stringent bounds, ranging from $\sim 10^{-7}$ to $\sim 10^{-10}$,
depending on the pseudoscalar mass~\cite{bazarko}.
\begin{figure}[ht]
\epsfxsize 3.3 in \epsfbox{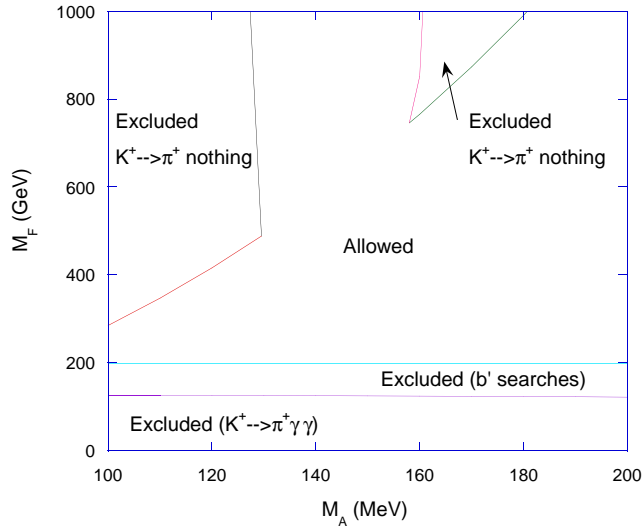}
\caption{Allowed parameter space for the fermiophobic scenario.}
\label{paramfig}
\end{figure}
In Fig.~\ref{paramfig} we display the allowed region of the model's
parameter space.  Within the two excluded regions toward the top of the
figure, the pseudoscalar is long lived enough to escape the detector, while
the branching fraction exceeds the bounds given in Ref.~\cite{bazarko}.
The gap between these regions corresponds to a mass interval in which there
are larger experimental backgrounds.  Immediately below each of these excluded regions, 
the pseudoscalar decays to two photons within the detector (assumed to have a 
fiducial length scale of $1.45$ meters~\cite{bazarko2}) and the weaker bound 
in Eq.~(\ref{kbound1}) becomes relevant.  However, one never reaches the region of 
parameter space excluded by the $K^+ \rightarrow \pi^+\gamma\gamma$ bound since 
the vector-like matter would itself become light enough to be detected in direct 
collider searches. We will restrict ourselves to the allowed regions of 
Fig.~\ref{paramfig} with smallest $M_F$ in discussing pseudoscalar production 
rates, in the next section.

Finally, we should comment on the bounds from the analogous decays of
neutral kaons and $B$ mesons.  First, the $K^0_s$ indeed may decay
into $\pi^0 \, A$; however, the total width of the $K^0_s$ is
approximately two orders of magnitude larger that that of the $K^+$, so
the branching fraction to the decay mode of interest is suppressed by
this factor relative to our previous results.  We therefore obtain no
further bounds.  The $K^0_L$, on the other hand, has a total width
that is about a factor of four smaller than that of the charged kaon.
However, the decay $K^0_L\rightarrow \pi^0 A$ is CP violating, so that
the decay amplitude is suppressed by an additional CP-violating spurion
factor of $\sim 10^{-3}$~\cite{goity}, and again no further bound is obtained. 
In the $B$ system, the decay $B\rightarrow K \eta'$ is observed, and
has a branching fraction of order $10^{-5}$~\cite{rpp}.  Using our previous result
for the $A \eta'$ mixing angle, we estimate that the branching fraction
for $B \rightarrow K A$ is ${\cal O}(10^{-15})$ and no further bound is
obtained.

\section{Production at Jefferson Lab}

We have seen that there is a window for light pseudoscalars in the
$100-200$ MeV mass range.  For the two-doublet Model II (or Model I
with $\tan\beta\sim 1$) the window requires substantial fine-tuning of
the charged Higgs mass; for the two-doublet Model I with large
$\tan\beta$, there is less fine-tuning, and for the fermiophobic case
there is a very large region of allowed parameter space.  How can one
detect these pseudoscalars?

A number of authors have considered light pseudoscalar detection at
high-energy colliders~\cite{cpyuan,others}.  Larios, Tavares-Velasco and 
Yuan~\cite{cpyuan} discussed production at the Tevatron, the LHC and 
future colliders.  They focused on the two-photon decay mode, which at 
high energies registers as a single photon signature.    In this
section, we consider the possibility of detecting the pseudoscalars
we have discussed in a beam dump experiment at the Thomas Jefferson 
National Accelerator Facility (Jefferson Lab).

Jefferson Lab has a high intensity photon beam directed into the CLAS
detector in Hall B.  The maximum energy is currently 6 GeV with an
upgrade to 12 GeV planned.  The photon beam has a bremstrahlung
spectrum with a luminosity of approximately $10^{34}\ {\rm cm}^{-2}\
{\rm sec}^{-1}$ if the photons are untagged.  At the 12 GeV upgrade a
monochromatic 9 GeV photon beam will also be available, with a
luminosity of approximately $10^{33}\ {\rm cm}^{-2}\ {\rm sec}^{-1}$.
The amplitude for pseudoscalar photoproduction may receive two
possible contributions. In the conventional two-Higgs-doublet models,
one can photoproduce the pseudoscalar most copiously off the strange
quark sea in the proton.  Second, in all the models we have discussed,
the pseudoscalar may bremstrahlung off the incident photon via the
loop-induced $A\gamma\gamma$ vertex.  Once produced, the pseudoscalar
will travel some distance and then decay into either $e^+e^-$ or
$\gamma\gamma$, depending on the model.  If the beam dump consists of
a meter or more of material, then most of the $\gamma\gamma$ background 
events will be suppressed.  It is thus important that the lifetime of the 
$A$ be sufficiently long that a substantial number make it through the beam
dump.

We first concentrate on production.  Consider photoproduction of the
pseudoscalar off the strange quark in the proton.  The parton level
cross section in the center of mass frame is
\begin{eqnarray}
{d\hat{\sigma}\over d\cos\theta}& =& {h^2e^2 p\over 144 \pi \hat{s}^{3/2}} 
\left[\frac{m_s^2-\hat{t}}{\hat{s}-m_s^2}+\frac{2 m_s^2 m_A^2}{(\hat{s}-m_s^2)^2}
+\frac{\hat{s}-m_s^2}{m_s^2-\hat{t}} \nonumber \right. \\ 
&+&\left.\frac{2m_s^2 m_A^2}{(m_s^2-\hat{t})^2}
+\frac{\hat{t}\hat{s}-(m_A^2+m_s^2)(\hat{s}+\hat{t}
)+m_s^4+m_A^4}{(\hat{s}-m_s^2)(m_s^2-\hat{t})} \right]  \,\,\, .
\end{eqnarray}
Here, $h$ is the Yukawa coupling of the $A$ to the strange quark, $p$ is the $A$ 
momentum; we have approximated the initial photon momentum as $\sqrt{\hat{s}}/2$
in the phase space factors to simplify the expression. In finding the full cross section 
for photoproduction, we multiply by the parton distribution function for the strange quark and 
integrate. However, since the parton model becomes less reliable at small
momentum transfers, one must keep in mind that there is significant
theoretical uncertainty from the small $x$ region of integration,
where the partonic cross section is largest. We therefore cut off the
$x$ integration at a value where $\hat{s}=xs= 1$ GeV$^2$.  We believe
that this choice is reasonable.  At lower $\hat{s}$ there will not be
enough energy to produce $\phi$, $\eta$ and $K$ mesons, and thus one
expects an additional suppression from the electromagnetic form factor 
due to the decrease in available exclusive channels. The resulting cross 
section is rather insensitive to the beam energy, varying from
$3.6\cot^2\beta$ to $2.0\cot^2\beta$ femtobarns as the photon energy
varies from $2$ to $12$ GeV.  For a luminosity of $10^{34}\ {\rm
cm}^{-2}\ {\rm sec}^{-1}$, this will yield approximately
$800\cot^2\beta$ events per year.  In order to be detected, these
pseudoscalars must travel through a beam dump.  The lifetime, as
discussed in Section II, gives a decay length for a $100$ MeV
pseudoscalar of $1.2\tan^2\beta$ centimeters times the relativistic
factor of $E/M_A$.  Consider a $6$ GeV beam and $\tan\beta=1$.  The
decay length is then $72$ centimeters, and roughly $25\%$ of the
particles, or $200$ particles/year, will travel through a one-meter
beam dump.  Since the differential cross section has a $t$-channel
pole in the massless quark limit, it is forward peaked and this
estimate will not suffer a substantial solid angle dilution.
As $\tan\beta$ increases, the production cross section
drops, but the decay length increases.  In Table II, we show the
number of events that traverse a one-meter beam dump per year,
assuming $10^{34}\ {\rm cm}^{-2}\ {\rm sec}^{-1}$ luminosity.  These
pseudoscalars will primarily decay into an electron-positron pair.
One should keep in mind that the uncertainties caused by the low $x$
cutoff could be substantial, and thus these event rates are approximate.  
Also, for larger beam energies and larger $\tan\beta$,
the decay length will be too long for a substantial number of events
to occur in a detector.  Nonetheless, the relatively high event rate
indicates that further experimental analysis is warranted.

\begin{table} [ht]
\center{
\begin{tabular}{cccccc} \hline\hline
$\tan \beta$ & $4\ {\rm GeV}$ & $6\ {\rm GeV}$ & $9\ {\rm GeV}$
& $12\ {\rm
GeV}$ & $24\ {\rm GeV}$ \\ \hline
1 & 125 & 210 & 350 & 340 & 340\\
2 & 150 & 147 & 144 & 135 & 120\\
3 & 88 & 85 & 77 & 66 & 46\\
4 & 53 & 53 & 48 & 40 & 32\\
5 & 41 & 34 & 28 & 25 & 21\\
\hline\hline
\end{tabular} }
\label{table2}
\caption{The number of pseudoscalars traversing at least one meter for various
values of the beam energy and $\tan\beta$ in the two-doublet model with
photoproduction off the strange quark sea.  We have assumed a luminosity of
$10^{34}\ {\rm cm}^{-2}\ {\rm sec}^{-1}$ and a pseudoscalar mass of $100$ MeV.
Most will decay into an electron-positron pair.}
\end{table}

\begin{figure}[ht]
\epsfxsize 3.3 in \epsfbox{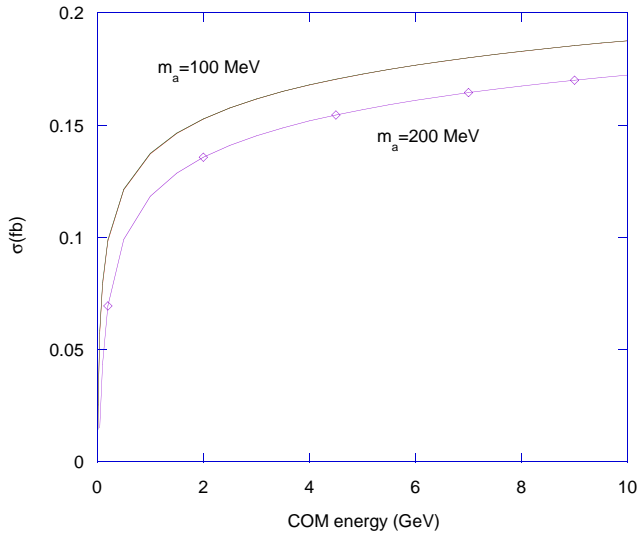}
\caption{Parton-level production photoproduction cross section in
the fermiophobic scenario as a function of center of mass energy, with
$M_F=200$~GeV and $\lambda=1$.}
\label{pro1}
\end{figure}

\begin{figure}[ht]
\epsfxsize 3.3 in \epsfbox{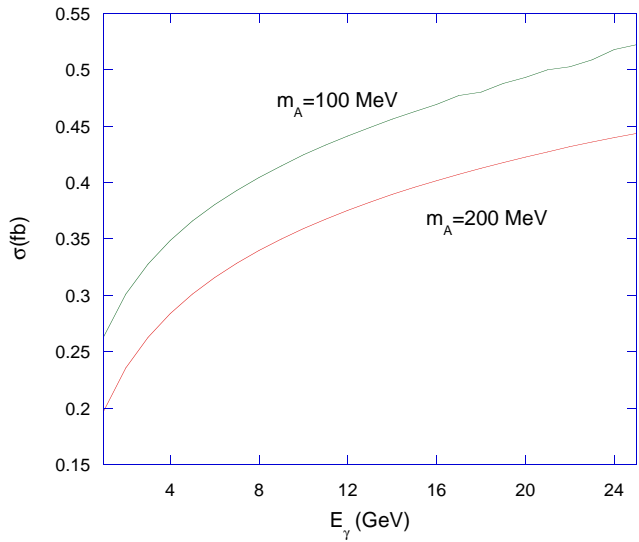}
\caption{Photoproduction cross section in the fermiophobic
scenario, as a function of photon beam energy in the lab frame,
with $M_F=200$~GeV and $\lambda=1$.}
\label{pro2}
\end{figure}

The second production mechanism is through the $A\gamma\gamma$ vertex.
In the two-Higgs-doublet models, the production mechanism already
considered strongly dominates, but in the fermiophobic model,
pseudoscalar bremstrahlung off the incident photon is the only
possibility. The parton level cross section is
\begin{equation}
{d\hat{\sigma}\over d\cos\theta}=-{\lambda^2Q^2e^6
p(
2m^2_qm^4_A+\hat{t}^3-2(m^2_A-\hat{s})\hat{t}^2+
((m^2_A-m^2_q)^2+m^4_q+2\hat{s}
^2-2(m^2_A+2m^2_q)\hat{s})\hat{t}
)\over
2048\pi^5M^2_F\hat{t}^2\hat{s}^{3/2}}
\end{equation}
where $\lambda$ is the coupling of the fermion in the loop to the $A$,
$M_F$ is the mass of the fermion in the loop, $p$ is the final state
3-momentum of the $A$, and $Q$ is the quark charge in units of $e$.  In 
deriving this expression, we have assumed that $M_F$ is much greater than the 
photon energy (certainly true for the fermiophobic case).  For $\sqrt{\hat{s}} >> m_q+m_A$,
we find that $\hat{\sigma}$ is well approximated by
\begin{equation}
\hat{\sigma} \approx \frac{\alpha^3 Q^2 \lambda^2}{64 \pi^2 M_F^2}\,
\frac{1}{\hat{s}}
\left[ (2\hat{s}-m_A^2)^2 \log \frac{(\hat{s}-m_A^2)^2}{m_q^2 m_A^2}-
3(\hat{s}-m_A^2)^2 \right] \,\,\, .
\label{eq:sghtapprox}
\end{equation}
The exact parton-level total cross section is shown in Fig.~\ref{pro1}.
The approximate expression given in Eq.~(\ref{eq:sghtapprox}) yields results
that are visually indistinguishable from those shown in Fig.~\ref{pro1}.
Using CTEQ set 5L structure functions for the up and down sea and valence
quarks we obtain the total production cross section shown in Fig.~\ref{pro2}. 
Assuming a monochromatic photon beam and a Jlab-like luminosity of
$10^{34}$~cm$^{-2}$ s$^{-1}$, one estimates 315 production events per year
per femtobarn of total cross section; qualitatively speaking, Fig.~\ref{pro2}
suggests ${\cal O}(10^2)$ events per year at an energy-upgraded Jlab, or
at some similar facility.

A more realistic analysis would take into account that the highest luminosity
photon beam at Jlab is not monoenergetic, but has a bremstrahlung spectrum.
We approximate this effect by assuming a total luminosity of
$10^{34}$~cm$^{-2}$ s$^{-1}$, with a distribution $dL/dE_\gamma \propto
1/E_\gamma$, with $E_\gamma$ ranging from $1$~GeV up to the beam energy;
the event rate is determined by the integral
\begin{equation}
\int \sigma \frac{dL}{dE_\gamma} d E_\gamma \,\,\,.
\label{eq:events}
\end{equation}
Table III shows the events per year for a number of different
choices for the beam energy and pseudoscalar mass.
\begin{table}[ht]
\begin{center}
\begin{tabular}{ccccc}
\hline\hline
$E_\gamma$ (GeV) &\hspace{1em}  & $m_A=100$~MeV & \hspace{1em} &

$m_A=200$~MeV\\\hline
6  && 99  && 79   \\
12 && 108 && 87 \\
24 && 117 && 96 \\ \hline\hline
\end{tabular}
\label{prodtable}
\caption{Photoprodution event rate per year in the fermiophobic scenario, with
$M_F=200$~GeV and $\lambda=1$. The total luminosity is taken to 
be $10^{34}$ cm$^{-2}$ s$^{-1}$ and a Bremstrahlung photon spectrum is assumed 
between $1$~GeV and the beam energy.}
\end{center}
\end{table}
Unlike the two-doublet model, the lifetime discussed in Section III is
sufficiently long that most of these pseudoscalars will traverse a
one-meter beam dump.  Another major difference is that these pseudoscalars
will decay into two photons, i.e. they will look like long-lived
$\pi^0$'s.    A more detailed analysis taking
into account possible experimental acceptances and cuts would be needed to
determine whether this signal could be separated from background under
realistic conditions.


\section{Conclusions}


We have considered light, elementary pseudoscalars with masses between
$100$ and $200$~MeV.  We have argued that such states may evade the
stringent bounds from $K$ and $B$ meson decays, while remaining of
interest in searches at low-energy photoproduction experiments, such
as those possible at Jefferson Lab.  In conventional two-Higgs doublet
models, light pseudoscalars may evade the strange and bottom meson
decay bounds due to a possible cancellation in the decay amplitude.
In this case, the coupling of the pseudoscalar to quarks is
substantial and one can produce the pseudoscalar state copiously via
photoproduction off the strange quark sea in a nucleon target.  On
other hand, if one wishes to avoid fine tuning in evading the decay
bounds, one can consider very natural scenarios in which the
pseudoscalar is fermiophobic. We have presented one concrete
realization of this idea, motivated by extra dimensions, and have
isolated the allowed parameter space of the model. In the fermiophobic
scenario, the pseudoscalar-two photon coupling leads to production via
pseudoscalar bremstrahlung off the incoming photon line.  The event
rate is substantial enough to make accelerator searches of potential
interest.




\begin{acknowledgments}
We thank Andrew Bazarko, Morton Eckhause, Keith Griffioen, Bohdan Grzadkowski,
Jon Urheim and C.P. Yuan for useful comments.  We thanks the National Science 
Foundation (NSF) for support under Grant No.\ PHY-9900657.  In addition, C.D.C. 
thanks the NSF for support under Grant Nos.\ PHY-0140012 and PHY-0243768.
\end{acknowledgments}


\end{document}